\documentclass[twocolumn,showpacs,preprintnumbers,amsmath,amssymb]{revtex4}

\usepackage{graphicx}
\usepackage{dcolumn}
\usepackage{bm}

\begin{document}

\preprint{adres preprintu}

\title{Shapiro effect for relativistic particles - testing General Relativity in a new window }

\author{Marek Kutschera}
 \altaffiliation[Also at ]{H. Niewodnicza\'nski Institute of
Nuclear Physics,Polish Academy of Sciences, Radzikowskiego 142, 31-342 Krak\'ow, Poland .}
\author{Wieslaw Zajiczek}%
 \email{wieslawzk@gmail.com}
\affiliation{%
M. Smoluchowski Institute of Physics, Jagellonian University, Reymonta 4, 30-059 Krak\'ow, Poland
}%

\date{\today}

\begin{abstract}
Propagation of relativistic particles in the Schwarzschild gravitational field  is studied. Particles emitted radially outward with speed at infinity exceeding  
$c/\sqrt{2}$ are observed to be accelerated
in  the gravitational field by a distant observer. This is the Shapiro effect
for relativistic particles. Slower particles are decelerated, as in Newtonian gravity. A speed-dependent potential for relativistic 
particles corresponding to the speed measured in terms of the coordinate time is derived
to be $V=GM\gamma(\gamma^2 -2)/r$ which is repulsive for $v > c/ \sqrt{2}$. The gravitational repulsion could be revealed in satellite 
experiments with
beams of relativistic particles subject to very precise time measurements. Principles of laboratory measurements able to test kinetic energy changes of relativistic particles in the 
Earth gravitational field are discussed.
 \verb+\pacs{#1}+ 
\end{abstract}

\pacs{Valid PACS appear here}
\maketitle

Gravitational interactions of light (electromagnetic radiation in general) have been studied extensively after prediction by A. Einstein \cite{einstein} of light deflection by massive 
bodies.
This \cite{eddington}  and other predictions of General Relativity, i.e. the gravitational redshift  \cite{pound} and the Shapiro  delay \cite{shapiro}  have been confirmed 
empirically. In contrast, little is known about
interactions of massive relativistic particles with the gravitational field. In this paper, a peculiar acceleration of relativistic particles receding from the central mass, which is a
Shapiro effect for relativistic particles, is studied. Experimental evidence of such an effect would provide a robust test
of General Relativity in a sector not tested yet.

Testing the interaction of relativistic with gravity would require using beams of relativistic particles and very  precise timing. Energy of particles should be such that the speed
range between $0.5c$ and $0.9c$ is covered. The physical nature of particles is irrelevant, however neutral beams might be easier to send over longer distances. 
 A crucial component of such experiments is very precise time measurement, possibly at subnanosecond level. Such a challenging requirement is a result of short time scale of the Shapiro delay,
which is the light crossing time of coordinate gravitational radius.

It was D. Hilbert who first noted \cite{hilbert22} that gravitational  interactions of relativistic particles in the Schwarzschild field are  different from those of nonrelativistic 
particles. 
From the Schwarzschild metric,
\begin{equation}
ds^2 = -\left(1-\frac{2m}{r}\right)dt^2 + \left(1-\frac{2m}{r}\right)^{-1}dr^2 + r^2d\Omega ^2,
\label{eq:metric}
\end{equation}
one can derive the radial geodesic equation
\begin{equation}
\frac{{\rm d}^2r}{{\rm d}t^2} = \frac{3m}{r^2}\left(1-\frac{2m}{r}\right)^{-1}\left(\frac{{\rm d}r}{{\rm d}t} \right)^2 - \frac{m}{r^2}\left(1-\frac{2m}{r}\right),
\label{eq:geodesic}
\end{equation}
whose integral is found to be
\begin{equation}
\left(\frac{dr}{dt}\right)^2 = \left(1-\frac{2m}{r}\right)^2-\left(1-\frac{2m}{r}\right)^3\left(1-v_0^2\right),
\label{eq:speed_coord}
\end{equation}
where $v_0$ is the particle speed at infinity.

The acceleration, Eq.~(\ref{eq:geodesic}),  strongly depends on the coordinate speed. If $v_0>1/\sqrt{3}$ - the particle 
is repulsed as ${{\rm d}^2r}/{{\rm d}t^2}>0$ for any value of $r>2m$ and the speed, Eq.~(\ref{eq:speed_coord}), monotonically grows. For $v_0<1/{\sqrt{3}}$, the 
speed, Eq.~(\ref{eq:speed_coord}), has a maximum at some finite value of $r=r_{max}$, and the acceleration is negative for $r>r_{max}$. For 
$v_0=0$, the case of free fall of a particle stationary at infinity, $r_{max}=6m$.

The Hilbert repulsion in Eq.~(\ref{eq:geodesic})  is a coordinate-dependent effect. There was a long-standing controversy as to the reality of this repulsion, for
 historical details see \cite{gruder}.

The repulsive effect of gravity which is a direct result of the geodesic equation, Eq.~(\ref{eq:geodesic}),  in the Schwarzschild metric was discovered by Shapiro \cite{shapiro}  
who considered a more invariant expression for change of the speed of light in the gravitational field \cite{shapiro1}. His prediction, which is now referred to as the Shapiro delay,
is  considered one of  classic tests of General Relativity.
 
The physical radial  speed, $v={dl}/{dt}$, is defined in terms of the metric, Eq.~(\ref{eq:metric}),   using the invariant lenght differential, 
$dl=dr/\sqrt{1-2m/r}$.
From the metric, Eq.~(\ref{eq:metric}), the speed of light measured in terms of the coordinate time is
\begin{equation}
\frac{dl}{dt}=\sqrt{1-\frac{2m}{r}}, 
\label{eq:speed_inv}
\end{equation}
and the acceleration is 
\begin{equation}
\frac{d^2l}{dt^2}=\frac{m}{r^2} \sqrt{1-\frac{2m}{r}}. 
\label{eq:accel1}
\end{equation}
Far from the central mass, $r>>2m$,  the acceleration is 
\begin{equation}
\frac{d^2l}{dt^2} \approx \frac{m}{r^2}=-g_N,
\label{eq:accel_far}
\end{equation}
where $g_N$ is the Newton acceleration. 

This formula shows that the speed of light measured by a distant observer is decelarated when light is falling onto the central mass, 
and accelerated when light is emitted from the central mass. The net result for the radio signal reflected by Venus in opposition  is a delay \cite{shapiro} with respect to the 
hypothetical signal 
travelling always with the constant speed of light c.

Jaffe and Shapiro \cite{jaffe72} noticed that also relativistic particles behave similarly as light in the gravitational field. 
They proposed an operational procedure to measure the speed of particles falling onto the central mass
which is independent of any coordinate system.  It consists of measuring the speed of a test particle using  the echo times of radio (light) signals 
reflected by the particle.  All time measurements refer to the coordinate time $t$ which is 
the proper time of a distant observer.

With the particle speed defined as for light, $v={dl}/{dt}$, using the invariant lenght differential $dl$, the invariant speed  is \cite{jaffe72}
\begin{equation}
\frac{dl}{dt} = \sqrt{\left(1-\frac{2m}{r}\right)-\left(1-\frac{2m}{r}\right)^2\left(1-v_0^2\right)},
\label{eq:speed_Jaffe}
\end{equation}
where $v_0$ is the speed at infinity. 
The acceleration measured by a distant observer is 

$$\frac{d^2l}{dt^2} = \frac{2m}{r^2}\left(1-\frac{2m}{r}\right)^{-1/2}\left(1+(v_0^2-1)\left(1-\frac{2m}{r}\right)\right)$$

\begin{equation}
-\frac{m}{r^2}\sqrt{1-\frac{2m}{r}}.
\label{eq:accel2}
\end{equation}

Far from the central mass, $r>>2m$, the acceleration of relativistic particles is  
\begin{equation}
\frac{d^2l}{dt^2} \approx \frac{m(2v_0^2-1)}{r^2},
\label{eq:accel11}
\end{equation}
which for $v_0=1$ is $-g_N$, the same as for light.
  
We conclude that relativistic particles moving out of the central mass  are observed by distant observer to be accelerated in the Schwarzschild field . 
The acceleration is almost exactly negative of the usual Newton attraction (deceleration) for ultrarelativistic particles with $v_0 \approx 1$.

One should stress  that the gravitational repulsion is only seen by a distant observer who measures speed at distant points with respect to his proper time. This was thoroughly 
discussed by McGruder \cite{gruder}. Local observers
would use 
the local time at their location, $T$, with $dT=(1-2m/r)dt$, to determine the acceleration. 
The speed of particles in terms of $l$ and time $T$ is
\begin{equation}
\frac{dl}{dT} =\sqrt{ 1-\left(1-\frac{2m}{r}\right)(1-v_0^2)}.
\label{eq:speed_loc}
\end{equation}
It is evident that this speed is a monotonically increasing function of $r$ and tends to $1$ for $v_0$ approaching the speed of light.
The acceleration is
\begin{equation}
\frac{d^2l}{dT^2} = \frac{m}{r^2}(v_0^2-1)\sqrt{1-\frac{2m}{r}} .
\label{eq:accel_loc}
\end{equation}
It is always negative an vanishes only in the limit $v_0=1$, i.e. for photons.
This is in  agreement with the result of Okun \cite{okun89}, who showed that particles whose speed approaches the speed of light  cease to be accelerated in the Newton
field.

The limiting speed, which for relativistic particles separates the regime of gravitational attraction from that of repulsion  can be obtained from 
Eq.~(\ref{eq:accel11}).
Gravitational repulsion starts to dominate for the speed $v_0>1/\sqrt{2}$.  

Comparing equations Eq.~(\ref{eq:speed_Jaffe}) and Eq.~(\ref{eq:speed_loc})  one concludes that the speed, Eq.~(\ref{eq:speed_Jaffe}),  for $v_0=1$ is the speed of light that 
is measured in Shapiro time delay experiment. The local speed, given by Eq.~(\ref{eq:speed_loc}),
is always equal to the speed of light $c$ (which is 1 in our units). Thus no delay is observed when the local time is used.

The Shapiro effect for relativistic particles could be verified in  experiments involving beams of relativistic particles with very precise timing. 
The time scale of the Shapiro delay is $\Delta \tau_S=r_g/c$, where $r_g=2m \equiv 2GM/c^2$ is the gravitational radius. For Earth $r_g=0.9cm$ and $\Delta \tau_S=3\times10^{-11}s$, a very
short time scale indeed. For sun
$\Delta \tau_S=10^{-5}s$, a significantly longer time scale, however sending beams of relativistic particles across the diameter of the Earth's orbit may  be difficult.

Consider an idealized version of the classical Shapiro experiment with a beam of relativistic particles sent along the radial direction from a place located at radial coordinate $r_1$ 
to another position at the same ray at $r_2$ where the beam is reflected
back to the receiver at $r_1$. 
The sender device, the receiver, and the reflector would be aboard satellites. If the satellite hosting both the sender of a beam of relativistic particles and the receiver, and a 
satellite carrying the reflector
are positioned at almost opposite points of  the same orbit (for the beam at most to 
graze the atmosphere) the classical Shapiro experiment could be performed. One should measure the time interval between sending and receiving the particles with speed $v_0$, $\tau(v_0)$,  
by precise determination of the  time instants of emitting  and receiving the particles.  
The formula, Eq.~(\ref{eq:speed_Jaffe}), of the Shapiro effect for relativistic particles then predicts that for beams sent with speed $v_0>1/\sqrt{2}$ the Shapiro time delay is measured: 
$\tau(v_0)>T=2d/v_0$, here $T$ is the time interval  needed for particles with speed $v_0$ to transverse twice the physical diameter $d$ of the orbit. 

For 
particles sent off with the speed $v_0<1/\sqrt{2}$ an opposite effect is measured: the time interval between sending and receiving back the
particles $\tau(v_0)$ is shorter than $T$. Such a time ordering is a common feature 
of Newtonian gravity,
when particles falling down in the gravitational field accelerate, and decelerate only when climbing up the potential well. 

Classical version of the Shapiro experiment requires time measurements by a single clock only, however problems with technical realization of a perfect reflection of relativistic particles 
without changing their speed could be an obstacle. Thus an experiment without reflector should be also considered. In this case a beam is sent from location $r_1$ to $r_2$ where a detector
coupled with precise clock measures the time instant of arrival of the beam. The two clocks, at the sending and receiving satellites should display the same time $t$ unaffected by gravity.
Also the clocks should be synchronized with sufficiently high precision 
presumably exceeding the present synchronization level of the GPS system clocks.

Relativistic particles with speed $v_0<1/\sqrt{2}$ ($\gamma_0<1.41$) behave in a Newtonian way in the gravitational field, with less relativistic particles being more Newtonian. 
Particles with speed $v_0=1/\sqrt{2}$ should not be delayed nor advanced, and thus should travel twice the orbit diameter in time $\tau(1/\sqrt{2})=T$ as if no gravity be present. 

Particles with speed  $v_0>1/\sqrt{2}$  start to behave in a "light-like" manner: the time to traverse the orbit diameter twice becomes longer than $T$, i.e. the Shapiro time delay occurs.
  The most delayed corpuscules are photons. 

The behaviour of fast particles is in a sense opposite to slow particles: such particles are deceletrated when falling down 
in the gravitational potential well, and accelerated only when climbing up the potential well. Relativistic particles  experience gravitational repulsion, a sort of apparent antigravity.

An urging question is what energy change of relativistic particles would be measured in an ideal laboratory experiment.

Let us consider  particles emitted outward in the Earth gravitational field, whose speed is measured using the coordinate time $t$. In the Newtonian limit
the kinetic and potential energy contributions to the particle energy can be defined. It is important that in the Newtonian limit time is assumed
not to be affected by gravitational field, and thus to advance with the same rate throughout the whole space-time. Thus the Newton time is a coordinate time
 
Formula, Eq.~(\ref{eq:accel2}), shows that both photons and relativistic  particles  are subject to gravitational repulsion. However, there is a crucial difference 
between photons and relativistic particles:  any acceleration bears no consequence for photon energies whereas a steady acceleration is expected to continuousely increase 
the kinetic energy of massive particles. 
Energy of a photon as a  constant of motion in the Schwarzschild metric is
\begin{equation}
E=h\nu(r)\sqrt{1-\frac{2m}{r}},
\label{eq:energy_phot}
\end{equation}
where $h\nu(r)$ is the energy measured by an observer stationary at $r$. The constant $E$ can be expressed in terms of the photon energy at infinity $E=h\nu_0$. It is clear that the
local photon energy, $h\nu(r)$, does not depend on the choice of time variable which determines how the speed of light is measured. In particular, it is not influenced by the acceleration,
 Eq.~(\ref{eq:accel1}), and remains the same as in the case of no acceleration, Eq.~(\ref{eq:accel_loc}).

For massive particles, the constant of motion $E$ in terms of the coordinate speed Eq.~(\ref{eq:speed_coord}) reads
\begin{equation}
E=\mu_0\sqrt{1-\frac{2m}{r}}\left(\sqrt{1-\left(\frac{dr}{dt}\right)^2\left(1-\frac{2m}{r}\right)^{-2}}\right)^{-1}.
\label{eq:energy_coord}
\end{equation}
Its value can be expressed by the Lorentz factor at infinity, $E= \mu_0\gamma_0$, where $\mu_0$ is the rest mass of a particle. In this formula, the speed is present explicitly 
which signals dependence on the time variable. This makes the real difference as comapred to photons in the weak field limit.      

For slow particles in the Newtonian limit, where gravity is very weak, one can divide the energy of test particles into kinetic and potential terms,  
$E=\mu_0(1+v^2/{2}+V_N(r))$, 
where $V_N(r)=-m/r$ is the Newton potential.

We derive a similar formula for relativistic particles using the formula Eq.~(\ref{eq:energy_coord}) and expanding the right-hand-side to lowest order in $2m/r$,
\begin{equation}
E=\mu_0\left(\sqrt{1-\left(\frac{dr}{dt}\right)^2}\right)^{-1} +2\mu_0\gamma_0\left(\gamma_0^2-\frac{3}{2}\right)\frac{m}{r}.
\label{eq:energy_split}
\end{equation}
Here the first term is the kinetic energy corresponding to speed measured in terms of the radial coordinate $r$ and the coordinate time $t$. The second term is the potential corresponding to 
such a choice of coordinates.

The conserved energy ,$E$, in terms of the invariant speed, Eq.~(\ref{eq:speed_Jaffe}), is
\begin{equation}
E=\mu_0\sqrt{1-\frac{2m}{r}}\left(\sqrt{1-\left(\frac{dl}{dt}\right)^2\left(1-\frac{2m}{r}\right)^{-1}}\right)^{-1}.
\label{eq:energy_inv}
\end{equation}
Expanding the rhs of Eq.~(\ref{eq:energy_inv}) to first order in gravitational coupling, the kinetic and potential terms can be identified as above.    
The kinetic energy is $\mu_0 \gamma(r)$, where $\gamma(r)=1/\sqrt{1-(dl/dt)^2}$ and   the potential is
\begin{equation}
V_{rel}(r,v_0)=\frac{m \gamma_0(\gamma^2_0-2)}{r}.
\label{eq:pot_relativistic}
\end{equation}

 For particles stationary at infinity, $v_0=0$, we find $V_{rel}(r,0)= -m/r$, the Newton potential. For $v_0=1/\sqrt{2}$ the potential vanishes, $V_{rel}(r,1/\sqrt{2})\equiv 0$.

For
relativistic particles with $v_0>1/ \sqrt{2}$, moving outward, the kinetic energy increases with $r$. In order for the energy to be conserved
the potential, $V_{rel}(r,v_0)$, should decrease with $r$, oppositely than the Newton potential does. 

The potential, Eq.~(\ref{eq:pot_relativistic}), is a generalizion of the Newton potential for relativistic particles corresponding to their speed being measured with respect to the coordinate 
time $t$. One thus can consistently describe the Shapiro delay of relativistic particles in weak gravitational fields as a result of the repulsive gravitational potential.

Let us consider principles of two different measurements of  the kinetic energy change of relativistic particles in the Earth gravitational field. In the first one, the kinetic energy is
measured by some time-of-flight method. To obtain particle speed at different levels one would prefer using clocks keeping the same time at every location. 
Current time keeping technology allows one to construct space endowed with a chosen coordinate time. In the GPS system, some coordinate time is maintained in a significant  volume of space 
surrounding Earth. One can imagine clocks in the whole volume to tick with the same rate in full synchronization, with gravitational influence eliminated by appropriate offsetting 
every clock before deployment. 

Relativistic particles with $v_0>1/\sqrt{2}$ emitted vertically from the Earth surface at $r_0$  display acceleration, when their speed is measured using the clocks maintaining the coordinate
time of the GPS sort.
According to the formula, Eq.~(\ref{eq:accel11}), the gravitational repulsion would be recorded. Using the time-of-flight method, the kinetic energy is obtained  from 
measured speed of particles. The 
particles at level $h$ ($h<<r_0$) above the ground are found to gain energy: $E_{kin}(r_0+h)=E_{kin}(r_0)+ \Delta E_{kin}$, where
$\Delta E_{kin}=V_{rel}(r_0,v_0)-V_{rel}(r_0+h,v_0)>0$.
Defining the redshift $z$  of kinetic energy as
\begin{equation}
z=-\frac{\Delta E_{kin}}{E_{kin,0}},
\label{eq:redshift}
\end{equation}
where $E_{kin,0}=\mu_0(\gamma_0-1)$ is the kinetic energy at infinity, we find 
\begin{equation}
z=gh \frac{\gamma_0(\gamma^2-2)}{\gamma_0-1}. 
\label{eq:kinetic1}
\end{equation}
Here $g=981 cm/s^2$ is the Earth's gravity.

For relativistic particles with $\gamma_0>\sqrt{2}$ redshift is negative, $z<0$, which actually corresponds to blueshift. 

One concludes that energy measurement of a time-of-flight kind together with use of the coordinate time would show that relativistic particles gain kinetic energy when moving vertically
upwards which indicates that gravity acts as a repulsive force.   

To check this unusual prediction  consider calorimetric measurement of kinetic energy of relativistic particles. We need to have definition of the physical kinetic energy of
particles, by which we mean a quantity independent of our choice of the coordinate time, uniquely determined by the physical time. It is commonly accepted that such physical time is 
the proper time $T$ of stationary observer at $r$.  

The splitting of the energy, Eq.~(\ref{eq:energy_inv}), into kinetic and potential parts is coordinate dependent, in particular it depends on the definition of speed. We can express 
the conserved energy in terms of the 
the local speed, $dl/dT$, given in   Eq.~(\ref{eq:speed_loc}). The energy is
\begin{equation}
E=\mu_0\sqrt{1-\frac{2m}{r}}\left(\sqrt{1-\left(\frac{dl}{dT}\right)^2}\right)^{-1},
\label{eq:energy_loc}
\end{equation}
and the corresponding potential obtained by expanding the rhs in $2m/r$ is
\begin{equation}
V_{loc}(r,v_0)=-\frac{m \gamma_0}{r}.
\label{eq:newton_loc}
\end{equation}
This is the usual attractive Newton potential for relativistic particles.

The kinetic energy change of relativistic particles emitted vertically from the surface after they are detected at high $h$ above the ground is equal to 
the difference of the potential energy Eq.~(\ref{eq:newton_loc})
at the two levels. The redshift $z$ is
\begin{equation}
z=gh \frac{\gamma_0}{\gamma_0-1}. 
\label{eq:kinetic}
\end{equation}
This redshift is positive for any velocity $v_0$.
 By definition, positive
values of redshift, $z>0$, correspond to the kinetic energy loss. 

For nonrelativistic particles with $\gamma_0 \approx 1$,  
the kinetic energy loss when climbing up the Earth gravitational field is the biggest: $z=gh/\epsilon$, where $\epsilon =\gamma_0-1 \to 0$. More relativistic particles loose smaller
fraction: for $\gamma_0=\sqrt{2}$ the redshift is $z=3.41gh$. Ultrarelativisric particles loose the same fraction of energy as photons.

The calorimetric measurement of the kinetic energy of relativistic particles  is expected to show the energy loss when particles move vertically upwards: the kinetic energy is 
redshifted in a similar way as for photons or nonrelativistic particles. This is the usual prediction of attractive gravity.

The astonishing difference of the two above discussed ideal measurements results from the very strong dependence of the gravitational interaction of relativistic particles on the time 
variable. This is not apparent for nonrelativistic 
particles. For particle velocities measured with respect to both the coordinate time and the physical time, the potential in the weak field limit is the same - the usual Newton potential. 
A common conclusion is that 
nonrelativistic dynamics is insensitive to gravitational time dilation in weak fields.

For relativistic particles the sensitivity to gravitational time dilation is extreme. Including the gravitational time dilation in the time variable defining velocities even in very 
weak fields changes the gravitational 
repulsion into attraction. This sensitivity could be employed to test the nature of the physical time underlying the dynamics of finite-rest-mass particles in 
gravitational fields.

 One could argue that  Eq.~(\ref{eq:kinetic}) is just the same as in Newtonian gravity for particle of inertial mass, $\mu_0\gamma_0$, in
the Newton potential, $V_N$, and thus no gravitational time dilation is needed and the universal Newton time (which is the coordinate time $t$) can be used to describe dynamics of
relativistic particles in very weak gravitational fields. However, for the Newton time kinematic acceleration of relativistic particles is measured according to Eq.~(\ref{eq:accel11}). 
Conservation of energy then requires that particles moving vertically upwards gain kinetic energy
Such particles are thus blueshifted, $z>0$, because the appropriate potential in this case is that of Eq.~(\ref{eq:pot_relativistic}). 

Neither of the predictions discussed above have  been tested yet. We focused only on principles of measurements able to test repulsion of relativistic particles.
The level of precision required in time, speed and energy measurements may not allow one to  carry out such experiments soon. However, when such precision is achieved, in addition to
testing the Shapiro effect for relativistic particles also the nature of physical time defining the kinetic energy of relativistic particles could be tested.  
Any discrepancy with the formula, Eq.~(\ref{eq:kinetic}), would indicate distortion of the physical time with respect to 
that assumed in Eq.~(\ref{eq:speed_loc}).

\bibliography{apssamp}

\end{document}